\documentclass[fleqn,usenatbib]{mnras}

\usepackage{newtxtext,newtxmath}
\usepackage[T1]{fontenc}

\DeclareRobustCommand{\VAN}[3]{#2}
\let\VANthebibliography\thebibliography
\def\thebibliography{\DeclareRobustCommand{\VAN}[3]{##3}\VANthebibliography}

\usepackage{graphicx}	% Including figure files

\usepackage{multirow}
\usepackage{dcolumn}
\newcolumntype{d}[1]{D{.}{.}{#1}}
\newcolumntype{v}[1]{D{|}{/}{#1}}

\usepackage[]{xcolor}

\definecolor{myg}{cmyk}{0.75002,0,1,0}
\newcommand{\myg}{\color{myg}}
\definecolor{msnote}{hsb:rgb}{0.692,0.692,0.692}

\definecolor{note}{cmyk}{0.0,0.5,0.6,0.}

\newcommand{\obj}{\mbox{3C~286}}

\newcommand{\feii}{Fe{\sc\,ii}}

\newcommand{\oiii}{[O{\sc\,iii}]$\lambda5007$}

\newcommand{\kmps}{$\rm km\,s^{-1}$}
\newcommand{\flux}{$\rm erg\,s^{-1}cm^{-2}$}

\newcommand{\kms}{km $\rm s^{-1}$}

\newcommand{\swift}{{\it Swift}}
\newcommand{\fermi}{{\it Fermi}}
\newcommand{\xmm}{{\it XMM-Newton}}
\newcommand{\chandra}{{\it Chandra}}

\title[Deep X-ray observations of 3C~286]{First deep X-ray observations of the \fermi-detected steep-spectrum source and radio-loud NLS1 galaxy 3C~286}

\author[S. Yao  et al.]{
Su Yao$^{1,2}$\thanks{E-mail: yaosu@bao.ac.cn},
S. Komossa$^{2}$\thanks{E-mail: astrokomossa@gmx.de}, 
A. Kraus$^{2}$, 
D. Grupe$^{3}$
\\
$^{1}$National Astronomical Observatories, Chinese Academy of Science, Beijing 100101, China\\
$^{2}$Max-Planck-Institut f\"ur Radioastronomie, Auf dem H{\"u}gel 69, 53121 Bonn, Germany\\
$^{3}$Department of Physics, Geology, and Engineering Technology, Northern Kentucky University, 1 Nunn Drive, Highland Heights, KY 41099, USA
}

\date{Accepted XXX. Received 2024; in original form}

\pubyear{2024}

\begin{document}
\label{firstpage}
\pagerange{\pageref{firstpage}--\pageref{lastpage}}
\maketitle

% Abstract of the paper
\begin{abstract}
A well-known calibrator source in radio astronomy, 
\obj\ ($z=0.85$), 
is a compact steep-spectrum (CSS) radio source and spectroscopically classified as a narrow-line Seyfert 1 (NLS1) galaxy. 
It is also known for its damped Ly$\alpha$ system from an intervening galaxy at $z=0.692$ detected in both ultraviolet (UV) and radio spectra. 
In addition, despite being a misaligned active galactic nuclei (AGN), \obj\ is also detected in $\gamma$-rays by \fermi. 
Thus, this unique object combines the characteristics of CSS sources, NLS1 galaxies, and $\gamma$-ray emitters with misaligned jets, providing an excellent laboratory for extending our knowledge of AGN disk-jet coupling. 
Despite its significance, \obj\ has been rarely observed in X-rays. 
In this study, we present our deep \xmm\ and \chandra\ observations of \obj. 
The results reveal that the X-ray spectrum can be well described by models including an intervening absorber with redshift and column density consistent with previous UV and radio observations. 
The most important finding is that the spectrum cannot be described by a single power law, 
but a soft excess is required which is parameterized by a blackbody.
Furthermore, we find evidence suggesting the presence of off-nuclear X-ray emission at a radius that corresponds to the location of the radio lobes. 
While further theoretical work is still needed, our findings offer new clues to understand the specific mechanism for $\gamma$-ray emission from this unique object. 
\end{abstract}

% Select between one and six entries from the list of approved keywords.
% Don't make up new ones.

%%%%%%%%%%%%%%%%%%%%%%%%%%%%%%%%%%%%%%%%%%%%%%%%%%

%%%%%%%%%%%%%%%%% BODY OF PAPER %%%%%%%%%%%%%%%%%%

\section{Introduction}\label{sec:introduction}

Active galactic nuclei (AGN) are among the most luminous long-lived sources in the Universe, 
and are suggested to be powered by their central supermassive black holes (SMBHs) with millions to billions solar masses through the process of accretion. 
Some of these accreting SMBHs in AGN are observed to generate jets, 
which can have a profound impact on the surrounding environment, 
including the evolution of galaxies. 
Despite the significant role of AGN jets in galaxy evolution, 
the mechanism behind their formation and their relationship with different accretion states of the SMBHs remain elusive.

Narrow-line Seyfert 1 galaxies and their high-luminosity equivalents among quasars (hereafter collectively referred to as NLS1 galaxies) make an exceptional subclass of AGN 
and are located at one extreme end of correlation space made up of AGN emission-line properties. 
NLS1 galaxies are conventionally identified by their Balmer emission lines from the broad-line region (BLR) with $\rm FWHM\lesssim2000$\,\kmps, relatively weak [O{\sc\,iii}] and strong \feii\ emission \citep[][]{1985ApJ...297..166O, 1992ApJS...80..109B}. 
They typically have steep X-ray spectra, strong outflow components, enhanced star formation activity, 
and are experiencing rapid growth of their central SMBHs 
\citep[review by][]{2008RMxAC..32...86K}. 
One of the special features of NLS1 galaxies is their radio emission.
In contrast to other AGN, NLS1 galaxies tend to be radio-quiet, with only 7 percent exhibiting radio-loud (RL) characteristics 
\citep{2006AJ....132..531K}.
A variety of studies on the RL NLS1 galaxies have suggested that those with flat radio spectra are actually compact steep spectrum (CSS) sources, 
which are thought to be young radio sources 
\citep[e.g.,][]{2001ApJ...558..578O, 2006MNRAS.370..245G,2006AJ....132..531K,2008ApJ...685..801Y,2016A&A...591A..98B}.
Further, we expect to see blazar-like objects among the flat-spectrum RL NLS1 galaxies. 
So far, more than 3000 classical blazars have been detected by \fermi/LAT in $\gamma$-rays 
\citep[][]{2020ApJS..247...33A}, 
meanwhile 
only less than 20 RL NLS1 galaxies have been detected in $\gamma$-rays 
(\citealt{2009ApJ...707L.142A}; Table~1 of \citealt{2018rnls.confE..15K}).
Considering their exceptionality among AGN and their characteristic difference from classical blazars, 
the study of these objects is important for our understanding of 
jet physics in a rarely explored parameter regime.

\obj, a well-known and widely-used calibrator source in radio astronomy, 
has been identified as an NLS1 galaxy (\citealt{2016PhDT.yao}; \citealt{2017FrASS...4....8B}; \citealt{2021MNRAS.501.1384Y}, hereafter paper I). 
It exhibits a broad H$\beta$ emission line with a full width at half maximum (FWMH) of approximately 2000\,\kms, 
a line flux ratio of \oiii/H$\beta<3$, 
and \feii/H$\beta>0.5$, 
meeting the criteria for classification as an NLS1 galaxies. 
\obj\ was detected in $\gamma$-rays 
with a significance of $\sim9\sigma$ and 
a Bayesian-based association probability of 98.5\% 
\citep[][]{2023ApJS..265...31A, 2023arXiv230712546B}. 
It is the radio-loudest one among all $\gamma$-ray-emitting NLS1 galaxies with a radio-loudness index that is even higher when compared to some broad-line radio galaxies (paper I). 
\obj\ was also identified as a CSS source \citep[][]{1982MNRAS.198..843P}. 
Its radio emission is compact on sub-arcsecond scales, with a jet, a counter-jet \citep[e.g.][]{2017MNRAS.466..952A}, 
as well as some outer emission that extends to $\sim3.8''$ 
\citep[][]{1989MNRAS.240..657S,1995A&AS..112..235A,2004ChJAA...4...28A}. 
The radio spectrum between 1.4 and 50\,GHz indicates a steep spectral index of $\alpha=-0.6$ ($S_{\nu}\propto\nu^{\alpha}$), 
and the viewing angle of its inner jet on tens of parsec scales was estimated to be 48$^{\circ}$ 
\citep[][]{2017MNRAS.466..952A}, 
indicating a mis-aligned jet (jet not pointing at us). 
Furthermore, 
\obj\ is also known for its damped Ly$\alpha$ system (DLA) from an intervening absorber at $z=0.692$, 
which has been detected in UV spectra 
\citep[][]{1992ApJ...399L.121M}
and in the H\,{\sc i} 21\,cm absorption 
\citep[][]{1973ApJ...184L...7B}, 
with a column density of $N_{\rm H}\sim2\times10^{21}\rm\,cm^{-2}$ 
\citep[][]{1994ApJ...421..453C}. 
The detection of diffuse emission 2.5 arcsec from \obj\ suggests 
that the intervening absorber is associated with a low surface brightness galaxy \citep[][]{1994AJ....108.2046S}.

Thus, in addition to its eminent relevance for radio astronomy as one of the few widely-used calibrator sources with very stable radio emission \citep[][]{2013ApJS..204...19P}, 
\obj\ is an exceptional object that combines the properties of CSS sources, NLS1 galaxies, mis-aligned $\gamma$-ray emitting AGN and DLA systems 
\citep[see Sect. 1 of][for a recent review of the multi-wavelength properties of 3C 286]{2024Univ...10..289K}. 
However, it has rarely been observed in X-rays. 
Previous observations taken by \chandra\ and \swift\ had exposures of only $\lesssim2$\,ks. 
%{\bf{
As a result, multi-component spectral fits could not yet be carried out. 
The X-ray spectra were found to be consistent with a simple single power law
with a photon index of $\Gamma\approx2$, 
and no excess of absorption was required beyond the expected Galactic contribution (paper I). 
This finding was inconsistent with the high column density observed in the DLA system and/or inconsistent with the SED modeling of \obj\ that requires an inverse Compton component in X-rays with $\Gamma\lesssim1.5$ \citep[][]{2020ApJ...899....2Z}. 
However, 
the statistics and/or resolution provided by these shallow observations are too limited to fit multi-component models with robust constraints on soft X-ray absorption, 
and they are also insufficient to detect any resolved X-ray emission in \obj. 
A detailed spectral and imaging analysis from deeper X-ray observations is required to unravel the puzzles about its high-energy emission.

Here we present the results of 
our X-ray observations of \obj\ using \xmm, \chandra, 
and quasi-simultaneous data from the Neil Gehrels \swift\ Observatory (\swift\ hereafter) and the Effelsberg radio telescope. 
This paper is structured as follows. 
Section~\ref{sec:obs_data} describes the observations and data reduction. 
The results and analysis are presented in Section~\ref{sec:results}. 
In Section~\ref{sec:discussion}, 
we discuss the nature and possible explanations of the results. 
Throughout this paper, 
we adopt a cosmology with 
$H_{\rm 0}$=70 km\,s$^{-1}$\,Mpc$^{-1}$, $\Omega_{\rm M}$=0.3 and $\Omega_{\rm \Lambda}$=0.7.

\section{Observations and data reduction}
\label{sec:obs_data}

\begin{table}
  \caption{Summary of observations used in this work.}
  \label{tab:obs_log}
  %\begin{tabular*}{\textwidth}{llclcr}
  \begin{tabular}{
        lr
        %v{+6}
        lcr}
  \hline
  \multicolumn{1}{c}{Telescope} & 
  \multicolumn{1}{c}{Observing Date} & 
  \multicolumn{1}{c}{Instrument (Band)} \\
  \hline
%  \noalign{\smallskip}
  \xmm & 2021-12-17 & EPIC (X-ray) \\ 
    &  & OM ($V$,$B$,$U$,$W1$,$M2$,$W2$) \\
  \chandra & 2023-02-12 & ACIS-S (X-ray) \\
  \swift & 2023-11-14 & UVOT ($V$,$B$,$U$,$W1$,$M2$,$W2$) \\
  Effelsberg & 2023-11-16 & ($\sim$2.6--50\,GHz) \\
  \hline
  \end{tabular}
\end{table}

\subsection{XMM-Newton}

We observed \obj\ on December 17, 2021, with the European Photon Imaging Camera 
(EPIC) 
%\citep[EPIC,][]{2001A&A...365L..18S,2001A&A...365L..27T} 
and the Optical Monitor (OM) 
%\citep[OM,][]{2001A&A...365L..36M} 
onboard \xmm\ (Table~\ref{tab:obs_log}). 
The data obtained from the observations were reduced following standard procedures using the Science Analysis System\footnote{\url{https://www.cosmos.esa.int/web/xmm-newton/what-is-sas}} 
({\sc sas version 21.0}) with the latest calibration files.

\subsubsection{EPIC}

The EPIC consists of three X-ray CCD cameras: 
one PN camera \citep[][]{2001A&A...365L..18S} 
and two MOS cameras \citep[][]{2001A&A...365L..27T}. 
They were operated in full frame mode with thin filter during the observation. 
The exposure times in PN, MOS1 and MOS2 are 42.8\,ks, 43.8\,ks and 43.9\,ks, respectively.
The data were reprocessed using {\sc sas} tasks {\it epproc} and {\it emproc} to generate calibrated event files. 
A high-energy light curve at $>10$\,keV was extracted from single events (PATTERN=0) across the whole field of view to identify time intervals of flaring particle background, 
that is, 
when the count rate is $>0.4$\,counts\,s$^{-1}$ for PN and $>0.35$\,counts\,s$^{-1}$ for MOS. 
The event files were then filtered and cleaned to remove flaring particle background and bad events. 
We also discarded events at $<0.2$\,keV and $>10$\,keV at this step. 
The resultant clean event files were used for the extraction of images, light curves and spectra of \obj.

Source photons were extracted from a circle with a radius of 40\arcsec\ centred on \obj. 
Background events were extracted from a source-free region. 
For the PN data, the background region was selected to be equidistant from the readout node as the source region, 
both located on the same chip. 
The light curves in 0.2--10\,keV were constructed with a bin size of 500\,s. 
The response matrix file (RMF) and ancillary response file (ARF) were generated using the tasks {\it rmfgen} and {\it arfgen}, respectively. 
The two MOS spectra were combined using the task {\it epicspeccombine}. 
The spectra were grouped such that each group contains data with a minimum signal-to-noise ratio of 5. 
The pile-up effect is negligible for the flux level of \obj\footnote{%
\url{https://xmm-tools.cosmos.esa.int/external/xmm_user_support/documentation/sas_usg/USG/epicpileup.html}}. 
The $\chi^{2}$ minimization technique was used during model-fitting in the following, 
as each group also contains photons larger than 25.

\subsubsection{OM}

The OM \citep[][]{2001A&A...365L..36M} was operated in imaging mode, 
utilizing all six filters during the observation. 
In each filter, 
multiple exposures were obtained. 
Standard aperture photometry was performed on the OM images, 
and the corrected count rates were converted to magnitudes and fluxes at the effective wavelengths of the filters using the OM image data processing chain, {\it omichain}. 

\subsection{Chandra}

We observed \obj\ with the Advanced CCD Imaging Spectrometer \citep[ACIS,][]{2002PASP..114....1W} onboard \chandra\ on February 12, 2023, 
for $\sim35$\,ks, 
which is significantly longer than the previous \chandra\ observation conducted in 2013. 
The chips S2, S3 and S4 were used, 
with \obj\ positioned on the aimpoint of the back-illuminated chip S3 to achieve optimal imaging and energy resolution\footnote{\label{cxc_chapter6}%
\url{https://cxc.harvard.edu/proposer/POG/html/chap6.html}}. 
The ACIS was operated in timed exposure mode, 
and the telemetry format was set to `Very Faint' mode. 
This setting offers the advantage of reduced background noise after ground processing$^{\ref{cxc_chapter6}}$, 
which would benefit the potential detection of faint structures.

The data were reprocessed using {\sc ciao (version 4.15)} 
with the latest calibration files. 
From the level 2 event file, 
which was created following standard procedures, 
full-resolution images and corresponding exposure maps were generated. 
These products were used to determine the appropriated extraction regions and to measure any structures. 
The position of \obj\ was determined using the program {\it wavdetect} included in the {\sc ciao} detect package. 
To construct the spectrum, 
we extracted source events from a circle centred on \obj\ with a radius of 4\arcsec. 
The background was determined in an annulus with inner and outer radii of 8\arcsec\ and 20\arcsec, respectively. 
The spectra, along with the corresponding RMF and ARF, were generated using the {\it specextract} script. 
The spectrum was grouped to ensure at least 25 counts in each group, 
and the $\chi^{2}$ minimization technique was used for model fitting.

\subsection{Swift}
\label{sec:uvot}

\obj\ was also observed with \swift\ \citep[][]{2004ApJ...611.1005G} on several occasions during the past few years 
(\citealt{2024Univ...10..289K}; paper I). 
Here, we use data from 
the UltraViolet and Optical Telescope \citep[UVOT,][]{2005SSRv..120...95R} onboard \swift,
which were obtained on November 14, 2023. 
This date is very close to the observing time of the Effelsberg radio telescope 
(Table~\ref{tab:obs_log}). 
The UVOT was operated in image mode with all six filters during the observations. 
Unfortunately, the data were taken when the \swift\ satellite underwent a small, uncorrected drift motion, 
causing all sources to appear elongated. 
Therefore, a larger source extraction region with a 15\arcsec\ radius was employed to ensure that the source photons are included 
(see \citealt{2024Univ...10..289K} for further discussion). 
The magnitudes in each filter were determined using the UVOT tool {\it uvotsource}, 
with the parameter {\it apercorr} set to `{\it curveofgrowth}' to compensate for the deviations from the calibrated circular source extraction region with a radius of 5\arcsec. %} 
The data were corrected for Galactic reddening using $E(B-V)=0.01$\,mag from \citet{2011ApJ...737..103S} and assuming the reddening curve in \citet{1989ApJ...345..245C}.

\subsection{Effelsberg radio observations}\label{sec:radiodata}

\obj\ serves as one of several calibrators 
in the OJ~287 MOMO project 
\citep[multi-wavelength observations and modelling of OJ~287, e.g.][] {Komossa2021, Komossa2023}. 
On November 16, 2023, we carried out radio observations (Table~\ref{tab:obs_log}) of \obj\ quasi-simultaneous with new \swift\ observations in the X-ray, UV and optical regimes. 
The goal was to reconfirm 
the constancy of the radio flux of \obj, as expected at the resolution of the Effelsberg telescope. 
Therefore, several other radio calibrators were employed (e.g. 3C~161, NGC~7027), 
and \obj\ was calibrated against these to measure its flux density.

The data were taken in the form of cross-scans, 
with several subscans in elevation and azimuth over the position of \obj. 
The amplitude (antenna temperature) and pointing offsets were determined by fitting a Gaussian profile to the individual subscans. 
The amplitudes of all subscans were averaged after correction for the typically small pointing offset. 
In the next step, 
corrections were made for the atmospheric attenuation as well as the telescope's gain-elevation effect (loss of sensitivity due to gravitational deformation of the main dish when tilted). 
Finally, the corrected antenna temperatures were compared with the expected flux densities and telescope sensitivities to convert the measurements into the Jansky scale.
Comparing with past radio measurements of \obj\ at similar angular resolution, 
we confirm that the flux densities are constant 
within the measurement accuracy.

\section{Results and Analysis}
\label{sec:results}

\subsection{Imaging}

\subsubsection{Source detection}
\label{sec:srcdet}

From the \chandra\ data, 
we created a full-resolution image in the 0.5--7\,keV energy range, 
an exposure map at a medium energy of 2.3\,keV, 
and a map of the point spread function (PSF) that encloses 90\% of the energy, 
using the script {\it fluximage}. 
Source detection was performed on the image using the program {\it wavdetect} included in the {\sc ciao} detect package. 
This yields the centre of \obj\ as 
RA=13$^{\rm{h}}$31$^{\rm{m}}$08\fs277, 
decl.=$+$30\degr30\arcmin33\farcs08, 
consistent with its optical position within 1\arcsec\ \citep[][]{2021A&A...649A...1G}.

\begin{figure*}
  %\vspace{-15pt}
  \centering
  \includegraphics[width=0.95\textwidth]{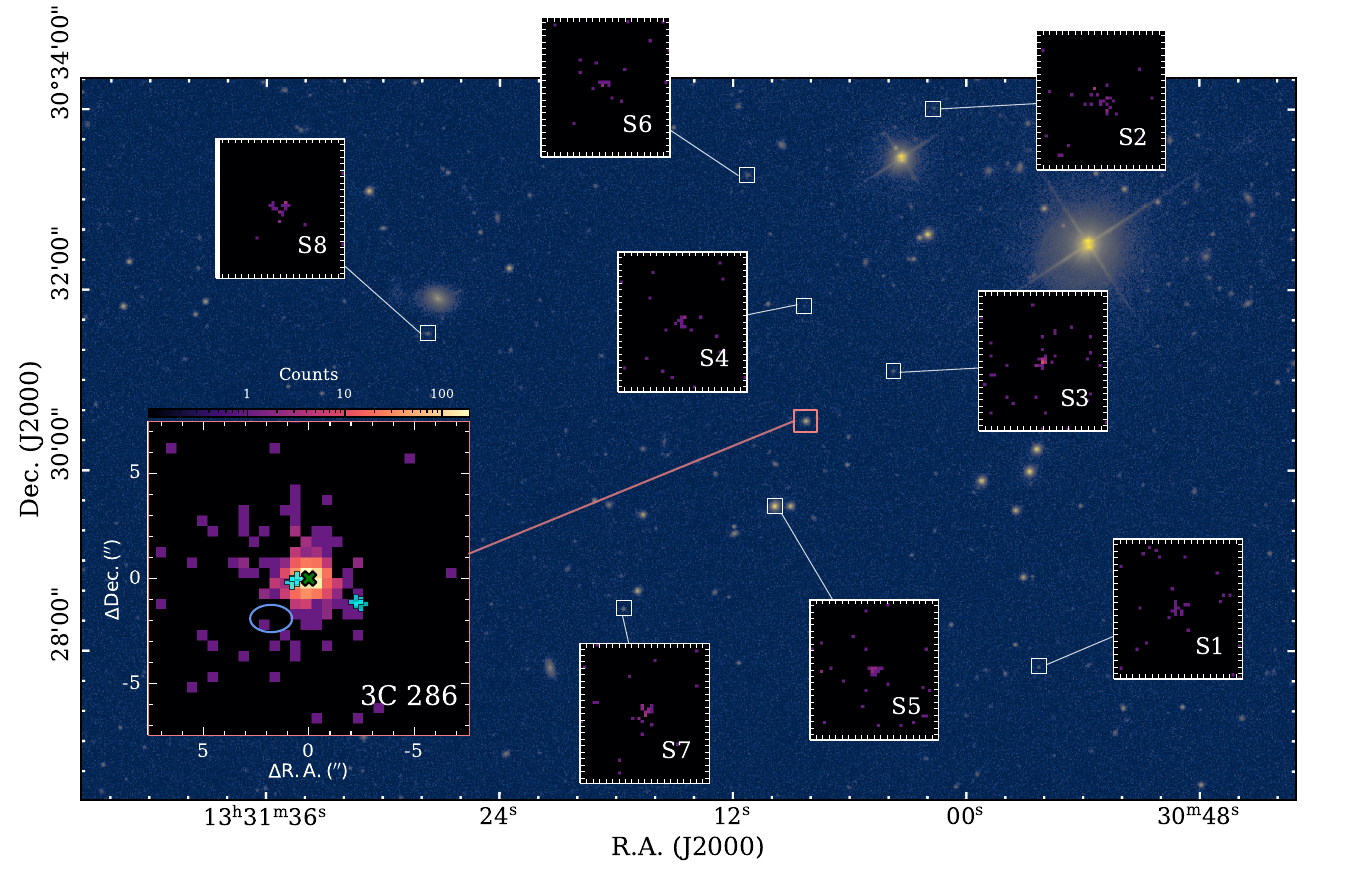}
%  \vspace{-8pt}
  \caption{
  SDSS $r$ band image of the \obj\ field, 
  with the inserted 0.5--7\,keV \chandra/ACIS X-ray images for both \obj\ and nearby X-ray sources that have optical counterparts (S1-S8). 
  While more X-ray sources are detected within this field, 
  they are not shown here and only listed in Table~\ref{tab:srcdet} (S9-S15).
  For \obj, the dark green ``X'' represents the center of its X-ray emission. 
  The extended components in the radio jet of \obj\ reported in \citet{2004ChJAA...4...28A}, E1/E2 and W1/W2, 
  are denoted by cyan ``$+$''. 
  The blue ellipse at $\sim$2.5\arcsec\ distance from \obj\ represents the location of the intervening faint galaxy at $z=0.692$ \citep[][]{1994AJ....108.2046S}. 
  }
  \label{fig:field_image}
%  \vspace{-10pt}
\end{figure*}

\begin{figure}
\centering
\includegraphics[width=0.9\columnwidth]{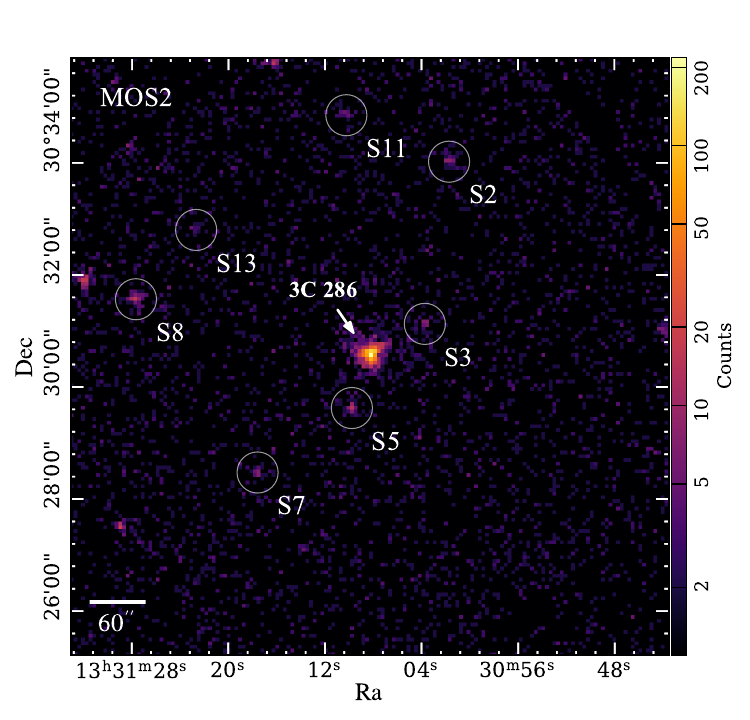}
\caption{%
MOS2 image centred on \obj. 
The nearby X-ray sources detected with both ACIS and MOS2 data are labeled in white circles. 
}
\label{fig:xmmmos}
\end{figure}

We detected 15 X-ray sources with a significance of 3$\sigma$ or higher within a $\sim$4-arcmin radius centered on \obj. 
These sources, along with their corresponding information, 
are listed in Table~\ref{tab:srcdet}. 
A search of the SDSS imaging photometric database for these X-ray sources revealed optical counterparts for 8 of them (S1-S8, Table~\ref{tab:srcdet}). 
Figure~\ref{fig:field_image} displays the positions of \obj\ and its nearby X-ray sources with optical counterparts on an SDSS $r$-band image, 
with their corresponding X-ray images inserted. 
Unfortunately, none of these nearby sources have optical spectroscopy data available. 
We have also run an \xmm\ source detection using the script {\it edetect\_chain} included in {\sc sas} following standard procedures\footnote{
\url{https://www.cosmos.esa.int/web/xmm-newton/sas-thread-src-find}
}, 
and detected the \chandra\ sources S2, S3, S5, S7, S8, S11 and S13 near \obj\ with \xmm\ as well (Figure~\ref{fig:xmmmos}).

\begin{table*}%[!h]
%\setlength{\tabcolsep}{4pt}
%\footnotesize
\caption{%
X-ray sources detected in the \obj\ field in the \chandra/ACIS image. 
%{\myg
S1-S8 are X-ray sources with optical counterparts shown in Figure~\ref{fig:field_image}, 
S9-S15 have no optical counterpart. %}
}\label{tab:srcdet}
\begin{center}             % used for centering table
\renewcommand{\arraystretch}{1.2}
\begin{tabular}{%
            c%{0.3\columnwidth}
            %D{,}{/}{-1} 
            ccc
            %D{,}{}{-1}
            ccccc
            %S[table-format=2]%{,}{}{-1}
            %S[table-format=1]
            }
    \hline
    \hline                 % inserts double horizontal lines
%    \multicolumn{1}{>{\centering}m{0.3\columnwidth}}{Source} & 
    \multicolumn{1}{c}{Name} & 
    \multicolumn{1}{c}{R.A.} &
    \multicolumn{1}{c}{decl.}  & 
    \multicolumn{1}{c}{Sig.$^{\dagger}$}  &
    \multicolumn{1}{c}{Count Rate$^{\ddagger}$}  &
    \multicolumn{3}{c}{SDSS Magnitude}  &
    \\%[0.1em]
    \multicolumn{1}{c}{} & 
    \multicolumn{2}{c}{(J2000)} &
    \multicolumn{1}{c}{} & 
    \multicolumn{1}{c}{($10^{-4}$\,ct/s)} &
    \multicolumn{1}{c}{$g$} &
    \multicolumn{1}{c}{$r$} &
    \multicolumn{1}{c}{$i$} &
    \\
    \hline
    \obj\ & 13:31:08.28 & +30:30:33.1 & 270 & 253.57$_{-15.26}^{+15.36}$ & 17.33 & 17.28 & 17.21 \\
    S1 & 13:30:56.27 & +30:27:50.3 & 4.4  & 3.79$_{-1.63}^{+2.24}$ & 22.55 & 21.42 & 21.12 \\
    S2 & 13:31:01.69 & +30:34:00.6 & 7.9  & 6.22$_{-2.13}^{+2.75}$ & 21.44 & 21.22 & 21.04 \\
    S3 & 13:31:03.75 & +30:31:06.5 & 12.0 & 7.10$_{-2.29}^{+2.90}$ & 21.17 & 20.94 & 20.41 \\
    S4 & 13:31:08.35 & +30:31:49.3 & 4.5  & 2.68$_{-1.31}^{+1.95}$ & 23.41 & 22.36 & 22.01 \\
    S5 & 13:31:09.82 & +30:29:36.5 & 6.0  & 3.35$_{-1.49}^{+2.12}$ & 16.62 & 15.19 & 14.05 \\
    S6 & 13:31:11.30 & +30:33:16.4 & 3.6  & 2.28$_{-1.21}^{+1.83}$ & 21.53 & 20.32 & 19.81 \\
    S7 & 13:31:17.62 & +30:28:28.6 & 8.0  & 5.98$_{-2.07}^{+2.66}$ & 20.06 & 20.12 & 19.84 \\
    S8 & 13:31:27.69 & +30:31:31.3 & 4.1 & 5.96$_{-2.07}^{+2.71}$ & 21.46 & 20.05 & 19.47 \\
    \hline %inserts single line
    S9 & 13:30:59.01 & +30:27:24.1 & 8.9 & 7.86$_{-2.43}^{+3.02}$ & $\dots$ & $\dots$ & $\dots$ \\
    S10 & 13:31:04.04 & +30:28:12.5 & 4.0 & 2.34$_{-1.22}^{+1.86}$ & $\dots$ & $\dots$ & $\dots$ \\
    S11 & 13:31:10.14 & +30:34:47.5 & 3.9 & 2.15$_{-1.19}^{+1.83}$ & $\dots$ & $\dots$ & $\dots$ \\% this is a inserted row, making index +1
    S12 & 13:31:12.26 & +30:32:06.5 & 3.3 & 1.94$_{-1.08}^{+1.71}$ & $\dots$ & $\dots$ & $\dots$ \\
    S13 & 13:31:22.85 & +30:32:45.4 & 5.9 & 3.80$_{-1.64}^{+2.27}$ & $\dots$ & $\dots$ & $\dots$ \\
    S14 & 13:31:25.43 & +30:31:24.7 & 4.3 & 3.83$_{-1.63}^{+2.24}$ & $\dots$ & $\dots$ & $\dots$ \\
    S15 & 13:31:26.37 & +30:30:02.7 & 4.1 & 3.05$_{-1.48}^{+2.12}$ & $\dots$ & $\dots$ & $\dots$ \\
    \hline 
\end{tabular}\\
\parbox[]{\textwidth}{%
%    {\bf Note. } \\%\footnotesize
    $^{\dagger}$ Source significance provided by {\it wavdetect}. \\
    $^{\ddagger}$ The net source count rate with 90\% confidence level uncertainties in the band 0.5--7\,keV in units of $10^{-4}$\,counts\,s$^{-1}$. 
    }
\end{center}
\end{table*}

\subsubsection{Source extension of \obj}

Previous radio observations have reported that \obj\ exhibits extended radio emission at kilo-parsec scales, 
displaying several structures within a few arcseconds around \obj\ \citep[][]{2004ChJAA...4...28A}. 
Taking advantage of \chandra's sharp PSF, 
it would be interesting to explore the potential presence of extended X-ray emission, 
which might provide clues to the radiative processes and properties of the jet in \obj. 

We obtained the radial profile of \obj\ from the \chandra\ image in the band 0.5--7\,keV using 12 equally-spaced annuli centered on the position of \obj. 
The annuli have a minimum radius of 0.5 pixels and a maximum radius of 12.5 pixels, respectively. 
The background was determined from an annulus with inner and outer radii of 25 and 50 pixels. 
As the PSF of \chandra\ varies greatly with position and energy, 
the best way to characterize its PSF is through simulations. 
We conducted ray tracing simulations using the software MARX \citep[][]{2012SPIE.8443E..1AD}, 
adopting the best-fit absorbed power-law spectral model for \obj\ (see Section~\ref{sec:specfit}) 
and spatially distributing it as a point source. 
The exposure time was set to correspond with  the observation, 
and the aspect solution from the observation was used to provide information on telescope dithering. 
The simulation was run 100 times to generate a combined PSF and a pseudo event file in the band 0.5--7\,keV. 
From these, we obtained the radial profile of the PSF and compared with that of \obj\ in order to explore the source's extension in X-rays.

\begin{figure}
\centering
\includegraphics[width=0.9\columnwidth]{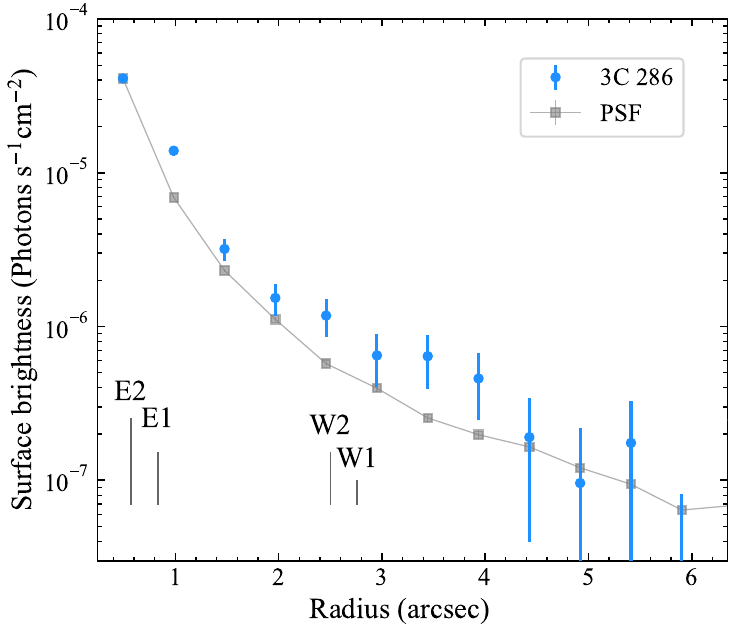}
\caption{
The radial profiles of \obj\ (blue) and the simulated point source (gray) generated by ray tracing simulations.
The PSF is scaled to match the peak of \obj. 
The distances from \obj's centre to the extended components `E1', `E2' at east and `W1', `W2' at west \citep[][]{2004ChJAA...4...28A} are also labeled. }
\label{fig:acis_psf}
\end{figure}

The radial profile of \obj\ and the simulated PSF are displayed in Figure~\ref{fig:acis_psf}. 
The excess of \obj\ over the PSF indicates the presence of extended X-ray emission beyond the wings of the PSF. 
Using the Very Large Array, 
\citet{2004ChJAA...4...28A} detected extended radio emission from components located 0.83 arcsec (E1) and 0.57 arcsec (E2) east of the radio core of \obj, 
as well as 2.76 arcsec (W1) and 2.5 arcsec (W2) to the west. 
We have labeled these radio components in both Figure~\ref{fig:field_image} and Figure~\ref{fig:acis_psf}. 
As can be seen, the excess X-ray emission is visible out to 4 arcsec from the core before fading into the PSF wings, 
albeit with uncertainties. 
This suggests that the excess X-ray emission likely originates from the same regions as the detected extended radio components (see also \citealt{2024Univ...10..289K}).

\subsection{Spectral fits}
\label{sec:specfit}

Spectral analysis of previous \chandra\ observation characterized the spectra of \obj\ with a single power law, 
having a photon index of $\Gamma\approx2$ and a Galactic absorption 
(paper I). 
We modeled our new \chandra\ spectrum in the band 0.3--10\,keV using a power law ({\tt pow}) and accounting for Galactic absorption, and found that this model fits the data well, 
yielding a photon index of $\Gamma=1.90\pm0.07$. 
Furthermore, the analysis of the \swift\ spectra as reported in \citet{2024Univ...10..289K} also demonstrated that a single power-law model can effectively describe the spectra. 
The photon indices, which exhibit large uncertainties, align with that of \chandra\ spectra. 
The \swift\ spectra are too shallow to warrant multi-component spectral fitting.

In addition to the Galactic absorption with a column density of $N_{\rm H}^{\rm Gal}=1.1\times10^{20}\rm\,cm^{-2}$ \citep[][]{2016A&A...594A.116H}, 
an intervening galaxy at $z=0.692$ was detected as an absorber 
with a column density of $\sim2\times10^{21}$ along the line of sight to \obj\ \citep[e.g.][]{1994ApJ...421..453C}. 
Consequently, additional absorption in soft X-rays is naturally expected. 
Furthermore, the presence of significant contributions from the accretion process as well as the relativistic jet to the SED (paper I) suggests that a more complex model than a single power law may be required to describe its X-ray emission. 
However, the spectra of \obj\ obtained by \chandra\ and \swift\ observations %
have insufficient statistics. 
A single power-law model was adequate to characterize the observed spectra well, 
yet it is challenging to extract reliable information regarding any additional absorption components within the model. 
With a larger effective area and a better response in the soft X-ray band, 
our \xmm\ observation provides better constraints on the absorption and other potential model components. 

\begin{figure}%[ht!]
  %\vspace{-15pt}
  \centering
  \includegraphics[width=0.99\columnwidth]{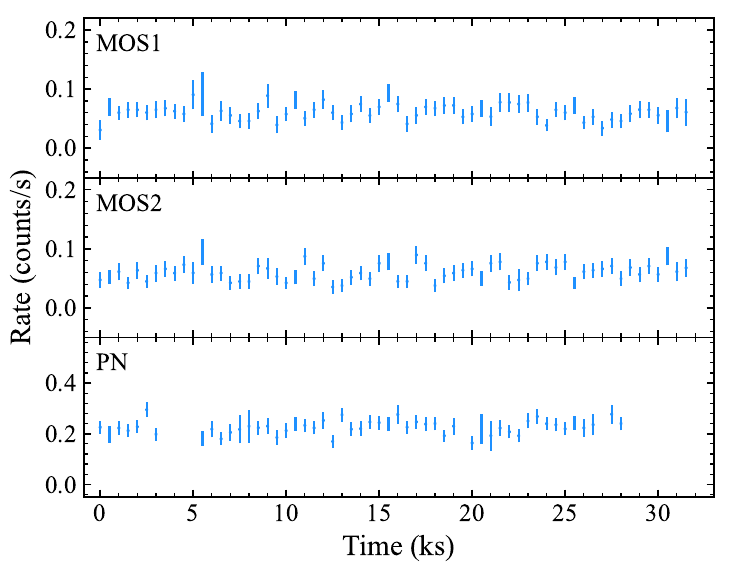}
%  \vspace{-8pt}
  \caption{
  The light curves of \obj\ obtained from PN, MOS1 and MOS2, respectively, in the energy range of 0.2--10\,keV. 
  }
  \label{fig:xmm_lc}
%  \vspace{-10pt}
\end{figure}

\begin{figure}%[ht!]
  %\vspace{-15pt}
  \centering
  \includegraphics[width=0.95\columnwidth]{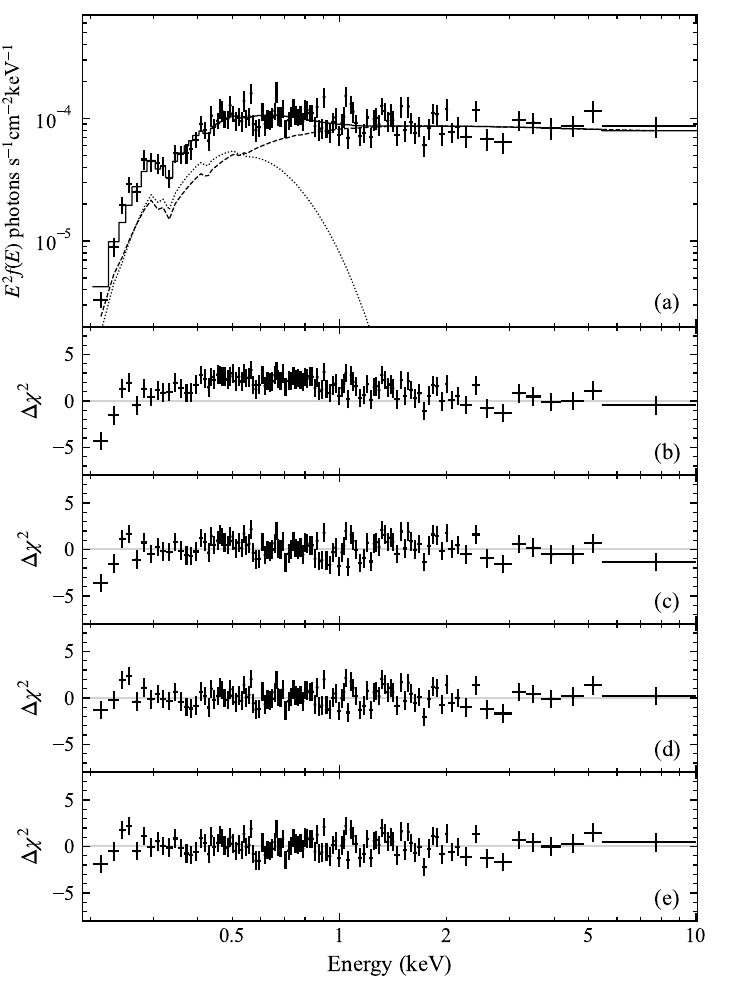}
%  \vspace{-8pt}
  \caption{
  Panel {\bf (a)} shows 
  the PN spectrum and the best-fit model of a blackbody ({\tt zbb}) plus power law ({\tt zpow}) absorbed by an intervening absorber with solar metallicity ({\tt tbvarabs}) and a Galactic absorber ({\tt tbabs}). 
  The blackbody and power-law component is represented by a dashed and dotted lines. 
  Panel {\bf (b)} shows 
  residuals of a single power law model fit to the 2--10\,keV data and extrapolated 
  to the soft X-ray band. 
  Systematic residuals are revealed $<$1\,keV. 
  The lower three panels show 
  residuals of the blackbody plus power-law model with Galactic neutral hydrogen absorption {\bf (c)}, blackbody plus power-law model absorbed by neutral hydrogen at $z=0$ but with column density free {\bf (d)}, 
  and blackbody plus power-law model absorbed by an intervening absorber with solar metallicity and a Galactic absorber. 
  }
  \label{fig:spec_fitting}
%  \vspace{-10pt}
\end{figure}

Figure~\ref{fig:xmm_lc} displays the light curves of \obj\ in the energy range of 0.2--10\,keV 
extracted from the PN and MOS detectors onboard \xmm. 
Upon inspecting the three light curves together, 
it is immediately obvious that there is no significant variability.
In addition, a variability test similar to the one conducted by \citet{2024arXiv240419107G} confirms that the source did not exhibit significant variability throughout the observation period.
Thus, we only consider the time-averaged spectra. 
We have focused our analysis on PN as it has a larger effective area than the two MOS combined, 
and provide more statistics especially at the soft band. 
A single power-law model ({\tt zpow}, $z=0.85$) with a local absorption fixed at the Galactic value yields a poor fit to the spectrum in the band 0.2--10\,keV (Table~\ref{tab:xmmspec}). 
It is commonly proposed that the hard X-ray continuum emission originates from the Comptonization of soft photons by a hot corona in the accretion process 
and/or by energetic electrons within a relativistic jet. 
In both scenarios, a power-law spectrum is expected in the narrow energy range of 2--10\,keV. 
So first we attempt to fit the PN spectrum in the band 2--10\,keV using a single power law, 
which is then extrapolated down to lower energies. 
As shown in Figure~\ref{fig:spec_fitting}b, 
the resulting residuals below $\sim$1 keV suggest the presence of an additional spectral component beyond a single power law. 
This indicates that a more sophisticated model is necessary to fully describe the spectra across the PN bandpass.

\begin{table*}%[!h]
\footnotesize
\caption{Results of spectral fits by a single powerlaw and blackbody plus power law ({\tt pow+zbb}) model with Galactic (Gal.) and intervening (Int.) absorptions.}\label{tab:xmmspec}
\begin{center}             % used for centering table
\renewcommand{\arraystretch}{1.4}
\begin{tabular}{%
            c%{0.3\columnwidth}
            D{,}{/}{-1} 
            cccccccccc
            %D{,}{}{-1}
            %S[table-format=2]%{,}{}{-1}
            %S[table-format=1]
            %c
            %c
            }
%            S%[table-format=4.2]
%            ccc}        % centered columns (4 columns)
    \hline
    \hline                 % inserts double horizontal lines
    \multicolumn{1}{c}{Absorption} & 
    \multicolumn{1}{c}{$N_{\rm H}^{\rm Gal}$/$N_{\rm H}^{\rm int}$} &
    \multicolumn{1}{c}{$kT$}  &
    \multicolumn{1}{c}{norm$_{\rm bb}$}  &
    \multicolumn{1}{c}{$\Gamma$}  & 
    \multicolumn{1}{c}{norm$_{\rm pl}$}  &
    \multicolumn{1}{c}{$\chi^{2}$/dof}
    \\%[0.1em]
    \multicolumn{1}{c}{Model} & 
    \multicolumn{1}{c}{[10$^{21}$\,cm$^{-2}$]} &
    \multicolumn{1}{c}{[eV]} &
    \multicolumn{1}{c}{[$10^{-6}$]} & & 
    \multicolumn{1}{c}{[$10^{-4}$]} &
    \\
    \hline
    \multicolumn{7}{c}{\tt zpow} \\
    Gal. & 0.11,  &  &  & $1.97^{+0.02}_{-0.02}$ & $3.05^{+0.06}_{-0.06}$ & 198.4/111 \\
    Gal.$^{\dagger}$ & 0.51^{+0.07}_{-0.06},  &  &  & $2.34^{+0.06}_{-0.06}$ & $4.71^{+0.31}_{-0.28}$ & 127.1/110 \\
    Gal. \& Int.$^{\ddagger}$ & 0.11, 1.03^{+0.15}_{-0.15} &  &  & $2.33^{+0.06}_{-0.06}$ & $4.58^{+0.28}_{-0.26}$ & 139.7/110 \\
    \multicolumn{7}{c}{\tt zpow+zbb} \\
    Gal. & 0.11,  & $319^{+14}_{-13}$ &  $4.95^{+0.60}_{-0.61}$ & $1.69\pm0.06$ & $1.70^{+0.18}_{-0.17}$ & 129.8/109 \\
    Gal.$^{\dagger}$ & 0.54^{+0.13}_{-0.11},  & $224^{+23}_{-22}$ & $5.90^{+1.82}_{-1.40}$ & $2.03\pm0.10$ & $3.09^{+0.46}_{-0.42}$ & 107.0/108 \\
    \multirow{3}{*}{Gal. \& Int.$^{\S}$} & 0.11, 1.75^{+0.62}_{-0.49} & $177^{+28}_{-23}$ & $11.32^{+9.22}_{-4.25}$ & $2.09\pm0.10$ & $3.40^{+0.49}_{-0.46}$ & 109.2/108 \\
        & 0.11, 2.94^{+1.03}_{-0.82} & $181^{+29}_{-24}$ & $9.72^{+7.19}_{-3.40}$ & $2.08^{+0.10}_{-0.09}$ & $3.34^{+0.47}_{-0.44}$ & 108.8/108 \\
        & 0.11, 6.47^{+2.19}_{-1.72} & $196^{+31}_{-29}$ & $5.94^{+3.01}_{-1.67}$ & $2.04\pm0.09$ & $3.15^{+0.41}_{-0.40}$ & 108.1/108 \\
    \hline
\end{tabular}
\parbox[]{\textwidth}{%
    {\bf Note. } norm$_{\rm bb}$ and norm$_{\rm pl}$ are normalisations of the blackbody and power-law components, respectively. \\
    $^{\dagger}$ The absorption within our Galaxy is given an initial value of $1.1\times10^{20}\rm\,cm^{-2}$ and set free during this model fitting. \\
    $^{\ddagger}$ The intervening absorber is at redshift of $z=0.692$ and set free during the fitting. \\
    $^{\S}$ The intervening absorber is set to have a solar, 50\% solar and 10\% solar metallicity, respectively. 
    }
\end{center}
\end{table*}

We begin by fitting the PN spectrum in the energy range of 0.2--10\,keV using phenomenological models. 
Both a broken power law 
and double power law are utilized to described the spectral shape, 
with Galactic absorption included. 
While the double power law model gives an unacceptable fit with $\chi^{2}=198.4$ for 109 degrees of freedom (dof), 
the broken power-law model provides a statistically good fit to the spectrum with $\chi^{2}/\text{dof}=112.5/109$, 
however, it yields a negative photon index of $-0.3$ below the break energy at $0.45\rm\,keV$, which is unphysical. 

Then we describe the spectrum using a single power-law model that is absorbed by an ionized absorber ({\tt absori})
%\citep[][]{1992ApJ...395..275D} 
at the redshift of \obj. 
The temperature of the ionized absorber is fixed at $3\times10^{4}\rm\,K$, 
with the column density and ionization left as free parameters. 
A local absorber ($z=0)$ is also included with hydrogen column density fixed at the Galactic value during the fitting. 
The best-fit result yields $\chi^{2}/\text{dof}=139.9/109$.

\begin{figure*}
  %\vspace{-15pt}
  \centering
  \begin{tabular}{lr}
		\includegraphics[width=0.95\columnwidth]{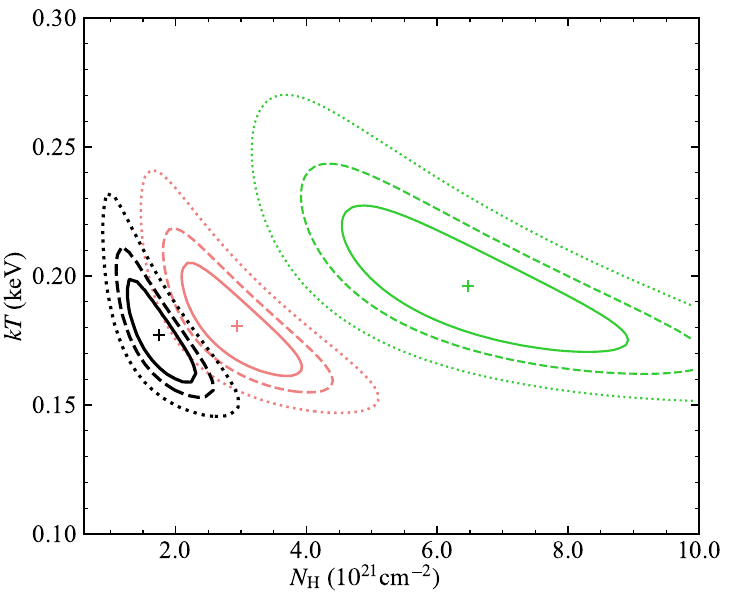} &
		\includegraphics[width=0.95\columnwidth]{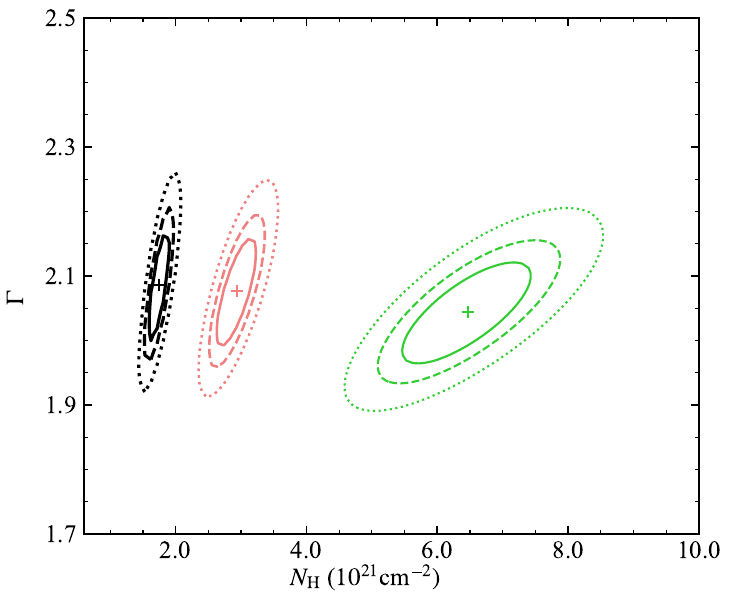} \\
	\end{tabular}
%  \vspace{-8pt}
  \caption{
  {\bf Left panel:} 
  Confidence contours at 68\% (solid), 90\% (dashed) and 99\% (dotted) for parameters $N_{\rm H}$-$kT$ derived from the model of a blackbody plus power law absorbed by the intervening ($z=0.692$) and Galactic absorbers, with the power-law component fixed at the best-fit values. 
  The black, red and green contours represent the results derived from the models having its intervening absorber with solar, 50\% solar and 10\% solar metal abundances. 
  The crosses represent the best-fit values in each case. 
  {\bf Right panel:} 
  Same as left panel but for parameters $N_{\rm H}$-$\Gamma$ with the blackbody component being fixed at the best-fit value. 
  }
  \label{fig:nh_contour}
%  \vspace{-10pt}
\end{figure*}

Next we employ a blackbody ({\tt zbb}, with a redshift of $z=0.850$) plus {\tt zpow} model, 
once again incorporating a local absorption fixed at the Galactic value. 
This model provides a much better fit ($\chi^{2}/\text{dof}=129.8/109$) compared to the single power law, 
as evidenced by the residuals displayed in Figure~\ref{fig:spec_fitting}c. 
Then we set the column density of the local absorption ($z=0$) as a free parameter in the {\tt zbb+pow} model and refit the data. 
This model provides an excellent fit with $\chi^{2}/\text{dof}=107.0/108$. 
However, 
the best-fit result yields a column density of $5.4\times10^{20}\rm\,cm^{-2}$, 
which is substantially higher than the Galactic value as measured by radio observations \citep{2016A&A...594A.116H}. 
This high column density suggests the presence of additional absorption beyond what is expected from neutral hydrogen in our own Galaxy. 
Taking into account the previous detection of an intervening galaxy at $z=0.692$ along the line of sight to \obj, 
we include an additional absorption component ({\tt tbvarabs}) at z=0.692, 
with its abundances fixed to solar values and its column density left as a free parameter. 
While the {\tt tbvarabs*zpow} model yields $\chi^{2}/\text{dof}=127.1/110$, 
the {\tt tbvarabs*(zbb+zpow)} provides a very good fit with $\chi^{2}/\text{dof}=109.2/108$, 
and yields a hydrogen column density of $N_{\rm H}^{\rm int}=1.75\times10^{21}\rm\,cm^{-2}$ for the intervening absorber. 
The flux in the band 0.3--10\,keV is $(4.9\pm0.2)\times10^{-13}$\,\flux\ after correction for Galactic absorption, 
and is $6.7^{+0.2}_{-0.3}\times10^{-13}$\,\flux\ after further correction for the intervening absorption. 
Remarkably, the derived column density of the intervening absorber closely matches the value of $2\times10^{21}\rm\,cm^{-2}$ measured from absorption lines in the optical/UV spectrum, 
reported in \citet{1994ApJ...421..453C}. 
However, measurements of the optical/UV absorption lines indicate that the metal abundances of the absorber are lower than solar. 
Therefore, we also test models with an absorption component having 50\% and 10\% of solar abundances, respectively. 
These models provide equally good fits but yield higher column densities, %
as shown in Figure~\ref{fig:nh_contour}. 
The best-fit model parameters are listed in Table~\ref{tab:xmmspec}.

\begin{figure}%[ht!]
  %\vspace{-15pt}
  \centering
  \includegraphics[width=0.95\columnwidth]{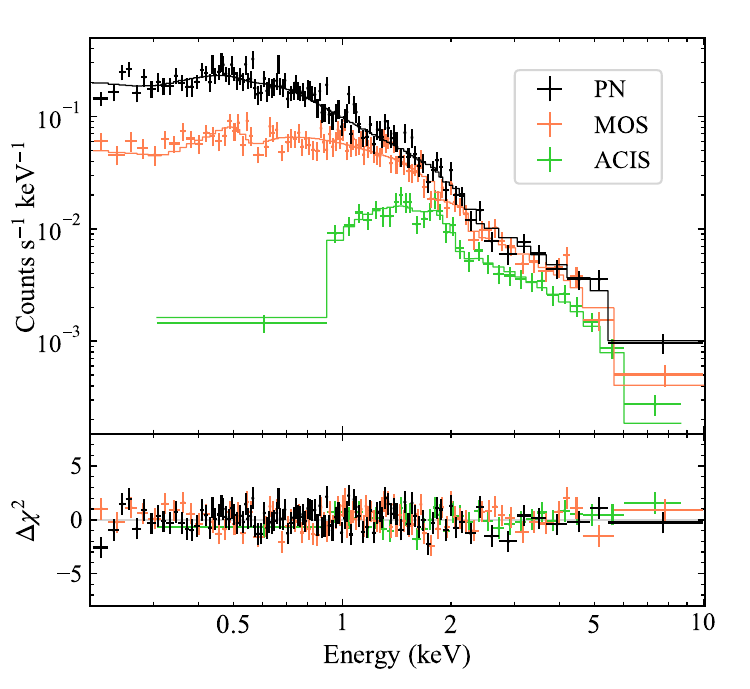}
%  \vspace{-8pt}
  \caption{
  The PN, combined MOS and ACIS spectra, 
  and the best-fit model (upper panel) consisting of a power-law component and a soft excess component parameterized by a blackbody, absorbed by Galactic and intervening absorber. 
  The residuals are shown in lower panel. 
  }
  \label{fig:joint_fit}
%  \vspace{-10pt}
\end{figure}

We note that we have checked that 
the PN spectral results are robust across various selections of source and background regions. 
We have also done a joint fit to the PN, combined MOS and ACIS spectra using the {\tt zpow+zbb} model with Galactic and intervening absorption at solar metalicity. 
Consequently, we obtain best-fit parameters comparable to those from the PN spectrum alone, 
with 
$N^{\rm int}_{\rm H}=1.19^{+0.40}_{-0.34}\times10^{21}\rm\,cm^{-2}$, 
$kT=193^{+24}_{-21}\rm\,eV$ and 
$\Gamma=1.98^{+0.06}_{-0.06}$. 
The best-fit model and spectra are displayed in Figure~\ref{fig:joint_fit}. 
Although the soft excess component is not required by the shallower \chandra\ data, 
when fitting the data sets jointly, 
the \chandra\ data is consistent with the presence of a soft excess component. 

\subsection{Broadband spectra}
\label{sec:sed}

\obj\ is very radio-loud and has been continuously detected in $\gamma$-rays, 
the actual reason for $\gamma$-ray emission is not yet known. 
In addition to non-thermal jet emission, 
\obj\ also exhibits signatures of thermal emission from the accretion process. 
Therefore, 
examining the broadband spectral energy distribution (SED) of \obj\ 
could provide valuable insights into the interaction between its accretion process and powerful relativistic jets. 
To construct the SED of \obj, 
we use simultaneous optical and UV data from \swift, along with quasi-simultaneous radio data collected with the Effelsberg telescope. 
The \swift/UVOT data of 2020 
(Paper I; \citealt{2024Univ...10..289K}) 
are also included as comparison. 
The UV and optical data have been corrected for Galactic redenning as mentioned in Section~\ref{sec:uvot}. 
The X-ray data are derived from \xmm\ PN observation, 
and have been corrected not only for Galactic absorption $N_{\rm H}^{\rm Gal}$, 
but also for intervening absorption $N_{\rm H}^{\rm int}=1.75\times10^{21}\rm\,cm^{-2}$ as detailed in Section~\ref{sec:specfit}. 
The $\gamma$-ray measurements between 100\,MeV--10\,GeV, 
as reported in the \fermi\ source catalog \citep[][]{2022ApJS..260...53A}, 
are incorporated. 
The data are presented in the observer's frame.

\begin{figure}%[t]
  %\vspace{-15pt}
  \centering
  \includegraphics[width=0.98\columnwidth]{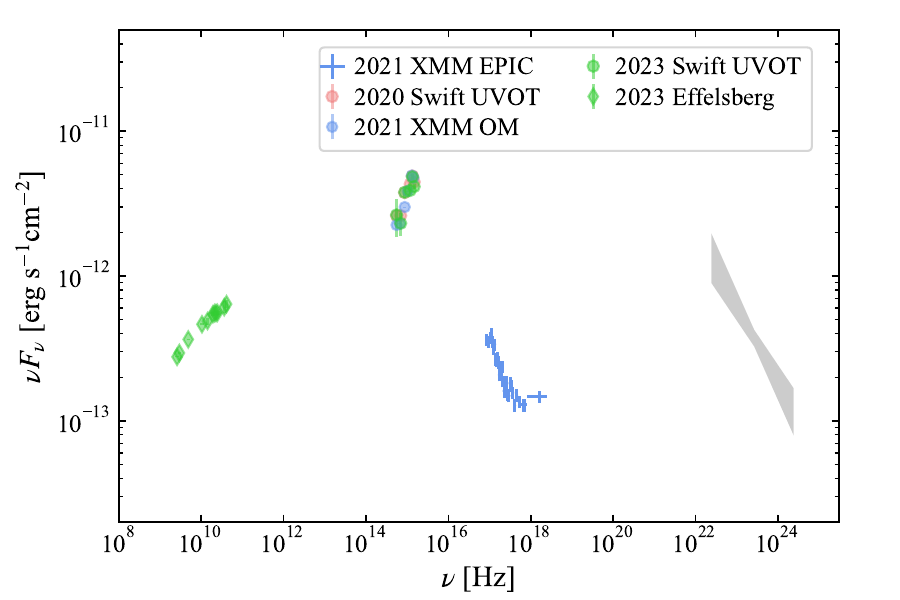}
  \vspace{-8pt}
  \caption{
  The SED of \obj\ in the observer's frame constructed with data from our Effelsberg, \swift\ UVOT, \xmm\ EPIC observations, 
  and the data from the \fermi\ point source catalog \citep[][]{2022ApJS..260...53A}. 
  The UV and optical data are corrected for Galactic reddening. 
  The X-ray data are corrected for Galactic absorption and intervening absorption obtained from the best-fit model (Section~\ref{sec:sed}) 
  and are re-binned for visual clarity.
  }
  \label{fig:sed}
%  \vspace{-10pt}
\end{figure}

\section{Discussion}
\label{sec:discussion}

We have presented the first deep imaging and spectroscopy observations of \obj\ in the X-ray regime, 
complemented by a new \swift\ observation and quasi-simultaneous radio data obtained from the Effelsberg telescope. 
These data sets are used to address some of the questions regarding the interpretation of the multiwavelength properties of \obj\, 
as highlighted in paper I.

\subsection{X-ray spectral shape, disk and jet contributions, X-ray absorption}

A suprising finding of paper I was the apparent lack of X-ray absorption, 
even though there were multiple lines of evidence for cold gas along the line of sight towards \obj\ \citep[e.g.][]{1994AJ....108.2046S}. 
However, these previous X-ray observations were very shallow, 
and did not provide constraints on multi-component model fits to the spectrum. 
This limitation has now been overcome with \xmm\ for the first time. 
In particular, it is well known that the presence of soft excess emission and additional absorption can compensate each other, 
potentially mimicking a single flat spectral component in data with low S/N and/or low spectral resolution. 
\xmm\ spectroscopy shows that a single power law no longer fits the data well. 
Once we account for an additional absorption component at redshift $z=0.692$, 
where its redshift and column density correspond to those inferred previously from HI 21cm absorption and the intervening galaxy observed as a DLA system 
\citep[][]{1973ApJ...184L...7B,1992ApJ...399L.121M,1994ApJ...421..453C}, 
we find that the spectrum is well described and requires the presence of an additional soft excess. 
Neither the soft excess nor the absorption were previously detected in X-rays. 
We interpret this soft emission component as the declining part of a `big blue bump', 
which we observe rising in the optical--UV spectrum with \swift, 
indicative of accretion-disc emission (Figure~\ref{fig:sed}). 
Several of the $\gamma$-ray detected NLS1 galaxies analyzed by \citet{2023MNRAS.523..441Y} have shown a blue bump extending into the X-ray regime, 
so \obj\ is not the only such source. 
The underlying hard X-ray power-law component could then represent the emission from a disc corona and/or jet.

\subsection{On the  origin of the $\gamma$-ray emission of 3C~286}

\obj\ is one of the few radio galaxies and radio steep-spectrum sources with a \fermi\ $\gamma$-ray detection. 
Radio galaxies are commonly suggested as `misaligned` blazars, 
which are observed at a larger angle between the line of sight and their jets 
\citep[][]{1995PASP..107..803U}. 
This perspective raises the question of the origin of the $\gamma$-ray emission from \obj.

A variety of different potential $\gamma$-ray emission models for \obj\ were briefly discussed by \citet{2024Univ...10..289K}. 
Given the unusual multi-wavelength properties of \obj, 
they considered the presence of multiple sources, such as a dual AGN, but have not yet found positive evidence for such a scenario. 
Microlensing has been ruled out, 
and there is no evidence to suggest that 3C 286 is a particularly young CSS source with the jet making its way out of the galaxy for the first time.

In paper I, 
we raised the question of whether the known intervening galaxy at redshift $z=0.692$ and $\sim$2.5\,arcsec offset from \obj\ could actually be an AGN and the counterpart to the $\gamma$-ray emission. 
However, this was rejected because the galaxy was not detected in the radio band.  
With our deeper \chandra\ observation, 
we have now searched for any X-ray emission from this galaxy. 
But we have not detected any isolated X-ray source associated with the intervening galaxy 
(Figure~\ref{fig:field_image}).

As the jet in \obj\ is not pointing directly at us, 
with a viewing angle of 48$^{\circ}$ \citep[][]{2017MNRAS.466..952A}, 
next we consider alternative scenarios for the $\gamma$-ray emission that have also been proposed for other misaligned jets detected in $\gamma$-rays. 
\citet{2008ApJ...680..911S} considered a scenario where  relativistic electrons, 
injected from hot spots into expanding lobes, 
could upscatter soft photons from the accretion disc or torus, 
producing high-energy emission that extends into the $\gamma$-ray regime. 
Alternatively, it has been also hypothesized that individual jet components might occasionally be 
deflected and then beamed toward the observer as they interact with dense blobs in the interstellar medium (ISM) along the jet's path 
\citep[][]{2012ApJ...755..170B, 2012A&A...539A..69B}. 
%(Barkov, Bosch-Ramon \& Aharonian 2012; Bosch-Ramon, Perucho \& Barkov 2012). 
In these scenarios, we expect enhanced off-nuclear X-ray emission, 
and have searched for any such emission around \obj. 
Although no isolated off-nuclear sources were detected, 
there is evidence for excess emission in the X-ray radial profile above the \chandra\ PSF, 
particularly at a radius where the extended radio components are located (Figure~\ref{fig:acis_psf}).

However, 
there remain challenges in understanding the $\gamma$-ray emission of \obj\ within these proposed scenarios. 
The $\gamma$-ray emission generated by the interaction between the dense region in the ISM and the jet along its path way is considered to be a temporary process \citep[e.g.][]{2012A&A...539A..69B}. 
\obj\ is regularly used as a calibrator in the radio regime, 
since it appears very stable with time.
In $\gamma$-rays, while \citet{2020ApJ...899....2Z} reported variability in \obj\ on timescales of hundreds of days, 
there is also evidence suggesting that \obj\ was stable in $\gamma$-rays during the past decade \citep[][]{2021MNRAS.507.4564P}. 
Further monitoring and timing analysis of \obj\ in X-rays and $\gamma$-rays might provide clues on whether the deflected jet scenario plays a role. 
But even so, 
in any case, the model focused particularly on young CSS sources, 
while \obj\ is already more evolved \citep{2024Univ...10..289K}.

In the scenario 
proposed by \citet{2008ApJ...680..911S}, 
high-energy emission is generated in the lobes through the upscattering of the soft photons. 
But, again, this scenario primarily focuses on young CSS sources rather than more evolved ones, such as \obj. 
Additionally, 
the X-ray flux level and spectral index depend on the energy of injected electrons, 
the size of jet, 
and the seed radiation sources for the inverse Compton scattering. 
These dependencies result in highly uncertain estimates of the X-ray emission during this process, 
making it challenging to test with current observations. 
Future deeper X-ray imaging may provide more precise measurements of 
the off-nuclear X-ray emission and offer constraints on the jet emission in \obj, 
if its $\gamma$-rays is really generated from the lobes.

\citet{2024Univ...10..289K} have noted that 
the near-Eddington accretion rate of the NLS1 galaxy \obj, 
and the reprocessing of disc photons at BLR and torus scales, 
generates a strong external photon field for efficient inverse Compton processes, 
in line with the finding of \citet{2020ApJ...899....2Z} 
that $\gamma$-ray detected CSS sources preferentially accrete at high Eddington rates.

\section*{Acknowledgements}
We would like to thank the \swift, \xmm, and \chandra\ teams for carrying out the observations we proposed. 
SY acknowledges support by an Alexander von Humboldt Foundation Fellowship between 2020 and  2022, when this project was started.
SK acknowledges the hospitality of NAOC Beijing in November 2023. 
This work is  partly based on observations obtained with the 100\,m telescope of the Max-Planck-Institut f\"ur Radioastronomie at Effelsberg. 
This research has made use of data obtained with the \chandra\ satellite, and software provided by the Chandra X-ray Center (CXC).
This work has made use of data obtained with \xmm, an ESA science mission
with instruments and contributions directly funded by ESA Member States and NASA. 
This research has also made use of data obtained from the High Energy Astrophysics Science Archive Research Center (HEASARC) provided by NASA’s Goddard Space Flight Center. 
This research has made use of the XRT Data Analysis Software (XRTDAS) developed under the responsibility of the ASI Science Data Center (ASDC), Italy. 
This work has also made use of the NASA Astrophysics Data System Abstract Service (ADS), and the NASA/IPAC Extragalactic Database (NED) which is operated by the Jet Propulsion Laboratory, California Institute of Technology, under contract with the National Aeronautics and Space Administration.

%%%%%%%%%%%%%%%%%%%%%%%%%%%%%%%%%%%%%%%%%%%%%%%%%%
\section*{Data Availability}

The data underlying this article are available in 
the astronomical archives of the HEASARC at
\url{https://heasarc.gsfc.nasa.gov/docs/archive.html},
the \swift\ archive at \url{https://swift.gsfc.nasa.gov/archive/}, 
the \chandra\ data archive at
\url{https://cda.harvard.edu/chaser/}, 
the \xmm\ data archive at \url{https://www.cosmos.esa.int/web/xmm-newton/xsa}, 
and will be shared on request.

%%%%%%%%%%%%%%%%%%%% REFERENCES %%%%%%%%%%%%%%%%%%

% The best way to enter references is to use BibTeX:

\bibliographystyle{mnras}
\bibliography{references}

% Alternatively you could enter them by hand, like this:
% This method is tedious and prone to error if you have lots of references
%\begin{thebibliography}{99}
%\bibitem[\protect\citeauthoryear{Author}{2012}]{Author2012}
%Author A.~N., 2013, Journal of Improbable Astronomy, 1, 1
%\bibitem[\protect\citeauthoryear{Others}{2013}]{Others2013}
%Others S., 2012, Journal of Interesting Stuff, 17, 198
%\end{thebibliography}

%%%%%%%%%%%%%%%%%%%%%%%%%%%%%%%%%%%%%%%%%%%%%%%%%%

%%%%%%%%%%%%%%%%% APPENDICES %%%%%%%%%%%%%%%%%%%%%

%\appendix

%\section{Some extra material}

%If you want to present additional material which would interrupt the flow of the main paper,
%it can be placed in an Appendix which appears after the list of references.

%%%%%%%%%%%%%%%%%%%%%%%%%%%%%%%%%%%%%%%%%%%%%%%%%%

% Don't change these lines
\bsp	% typesetting comment
\label{lastpage}
\end{document}